\documentclass[a4paper,11pt]{amsart}
\usepackage[utf8]{inputenc}
\usepackage{amsaddr}
\usepackage{calrsfs}
\usepackage{graphicx,color}
\usepackage[english]{babel}
\usepackage{graphicx}
\usepackage{amsmath}
\usepackage{epstopdf}
\usepackage{float}
\usepackage{color}

\usepackage{amssymb}

\numberwithin{equation}{section}

\theoremstyle{definition}

\theoremstyle{plain}
\newtheorem{thm}{Theorem}[section]

\newtheorem{cor}{Corollary}[section]

\theoremstyle{definition}

\begin{document}
\title[Fractional delta hedging strategy]
{Fractional delta hedging strategy for pricing currency options with transaction costs}

\date{\today}

\author[Shokrollahi]{Foad Shokrollahi}
\address{Department of Mathematics and Statistics, University of Vaasa, P.O. Box 700, FIN-65101 Vaasa, FINLAND}
\email{foad.shokrollahi@uva.fi}

\begin{abstract}
This study deals with the problem of pricing European currency options in discrete time setting, whose prices follow the fractional Black Scholes model with transaction costs. Both the pricing formula and the fractional partial
differential equation for European call currency options are obtained by applying the delta-hedging strategy. Some Greeks and the
estimator of volatility are also provided. The empirical studies and the simulation findings show that the fractional Black Scholes with transaction costs is a satisfactory model.
\end{abstract}


\maketitle
\section{Introduction}\label{sec:1}

A currency option is a contract that gives the holder the right to buy or sell a
certain amount of foreign currency at a fixed exchange rate (exercise price) upon
exercise of the option. There are two types of currency options: American options are options that can be exercised at any
time before they expire, while European options can be exercised only during a specified
period immediately before expiration.

Option pricing was introduced by Black-Scholes \cite{black} in 1973. Duan
and Wei \cite{duan1999pricing} indicated that option pricing by Black-Scholes
model which is based on Brownian motion cannot illustrate
clearly two phenomena from stock markets: first asymmetric
leptokurtic features and second the volatility smile. In a
work by Garman and Kohlhagen $(G-K)$ \cite{garman1983foreign} was
extended the Black-Scholes model in order to make valuation
European currency options, having two fundamental features:
(1) estimating the market volatility of an underlying asset
generally as a function of price and time without direct reference
to the specific investor characteristics such as expected yield, risk aversion measures, or utility functions; (2) self replicating
strategy or hedging. However, some researchers
(see \cite{cookson1992models}) presented some evidence of the mispriced currency
options by the $G-K$ model. The significant causes of why
this model is not suitable for stock markets are that the
currencies are different from the stocks in main respects and
geometric Brownian motion cannot resolve the conduct of
currency return, see \cite{ekvall1997currency}. Since then, in order to overcome
these problems, many systems for pricing currency options
were proposed by using amendments of the $G-K$ model \cite{rosenberg1998pricing,sarwar2000empirical,bollen2003performance}. Moreover, the empirical studies
also demonstrated that the distributions of the logarithmic
returns in the asset market generally reveal excess kurtosis.
It can be said that the properties of
financial return series are nonnormal, nonindependent, and
nonlinear, self-similar, with heavy tails, in both autocorrelations
and cross-correlations, and volatility clustering \cite{huang1995fractal,cajueiro2007long,ding1993long,podobnik2008detrended}. Since fractional Brownian motion $(FBM)$ has two important
properties called self-similarity and long-range dependence, it has the
ability to capture the typical tail behavior of stock prices or indexes \cite{wang2010scaling,wang2010scaling0,shokrollahi1,shokrollahi5,shokrollahi4}.

In classical finance theory, absence of arbitrage is one of the most unifying concepts. However, behavioral finance and
econophysics as well as empirical studies sometime propose models for asset price that are not consistent with this basic
assumption. A case is the fractional Black-Scholes $(FBS)$ model, which displays the long-range dependence observed in empirical
data \cite{ozdemir,mandelbrot1,mariani}. The $FBS$ model is a generalization of the Black-Scholes model, which is based on replacing the standard Brownian motion by a $FBM$ in the Black-Scholes model. Since $FBM$ is not a semimartingale \cite{liptser}, it has been shown that the $FBS$ model admits arbitrage in a complete and frictionless market \cite{cheridito,rogers,salopek,wang2010scaling0,shokrollahi2}. The purpose of this paper is to resolve this contradiction between classical Black-Scholes-Merton theory and practice through both giving up the arbitrage argument used by Black and Scholes to price currency options and examining option replication in the presence of proportional transaction costs in a discrete time setting. Moreover, we show that the time scaling and long-range dependence in return series exactly have an impact on currency options pricing whether proportional transaction costs are considered or not.

Leland \cite{leland} was the first who examined option replication in the presence of transaction costs in a discrete time setting.
From the point of view of Leland \cite{leland}, in a model where transaction costs are incurred at every time the stock or the bond is
traded, the arbitrage-free argument used by Black and Scholes [23] no longer applies. The problem is that due to the infinite
variation of the geometric Brownian motion, perfect replication incurs an infinite amount of transaction costs. Hence, he
suggested a delta hedge strategy incorporating transaction costs based on revision at a discrete number of times. Transaction
costs lead to the failure of the no arbitrage principle and the continuous time trade in general: instead of no arbitrage, the
principle of hedge pricing - according to which the price of an option is defined as the minimum level of initial wealth
needed to hedge the option - comes to the fore.

The rest of this work is as follows: some propositions and definitions are presented in Section \ref{sec:2}. We propose a new framework for pricing call currency options in discrete time setting by applying delta-hedging strategy and $FBS$ with transaction costs, in Section \ref{sec:3}. Furthermore, the impact of time-step $\delta t$ and Hurst parameter $H$ on our pricing model are discussed, in Section \ref{sec:3}. Section \ref{sec:4} deals with the simulation studies for our pricing formula, estimation of the volatility, and  the Hurst parameter $H$ for currency call option data from  China Merchants Bank $(CMB)$. Moreover, the comparison of our $FBS$ model with transaction costs and traditional models is undertaken in this Section. Section \ref{sec:5} is assigned to conclusion.

\section{Preliminaries }\label{sec:2}
In this section, we present some essential assumptions and definitions that we will need for the rest of the paper.
A  $FBM$, $B_H(t)$ with Hurst parameter $H\in (0,1)$ under  the probability space $(\Omega,F,P)$, is a continuous Gaussian process  with the following properties:
\begin{enumerate}
\item[(i)] $B_H(0)=0$
\item[(ii)]  $E[B_H(t)]=0$ for all $t\geq0$,
\item[(iii)]  $Cov[B_H(t)B_H(s)]=\frac{1}{2}\Big[t^{2H}+s^{2H}-|t-s|^{2H}\Big]$ for all $s,t\geq0$,
\end{enumerate}
If $H=\frac{1}{2}$, then the corresponding $FBM$ is the usual standard Brownian motion. It can be easily
seen that $E(B_H(t)-B_H(s))^2=|t-s|^{2H}$. Furthermore, $B_H(t)$ has stationary increments and is $H$-self-similar. More details
about the $FBM$ can be found in the paper \cite{lewellen}.

If $H>\frac{1}{2}$, the process $(B_H(t),\,t\geq 0)$ exhibits a long-range dependence, that is, if $r(n)=E[B_H(1)(B_(n+1)-B_H(n))]$,
then $\sum_{n=1}^\infty r(n)=\infty$. As mentioned in  \cite{chen}, long-range dependence is widespread in economics and finance and has
remained a topic of active research \cite{mandelbrot,cajueiro,cajueiro1}. Long-range dependence seems also an important
feature that explains the well-documented evidence of volatility persistence and momentum effects \cite{lewellen,cajueiro1}. Hereafter we
shall only consider the case $H\in(\frac{1}{2},1)$, which is most frequently encountered in the real financial data.

The groundwork of modeling the effects of transaction costs was done by Leland \cite{leland}. He adopted the hedging strategy of
rehedging at every time-step, $\delta t$. That is, every $\delta t$ the portfolio is rebalanced, whether or not this is optimal in any sense. In the following proportional transaction cost currency options pricing model, we follow the other usual assumptions in the Black-scholes
model but with the following exceptions:

\begin{enumerate}

\item[(i)] The portfolio is reviewed in each finite, constant and small interval $\delta t$.

\item[(ii)] Transaction costs are proportional to the value of the dealing in the financial assets. Assume that $U$ contributions are purchased $(U>0)$ or sold $(U<0)$ at the value $S_t$, hence the trading costs are defined as $\frac{\alpha}{2}|U|S_t$  in both cases of purchasing and selling. Furthermore, trading occurs just at interval. In the $FBS$ model, the trading of stocks or the bonds has transaction costs in any interval of times, the no-arbitrage strategy utilized just by Black and Scholes. Infinite variation is considered as an obstacle in the geometric $FBM$, and in the  unlimited value of dealing costs due to total replication.
\item[(iii)] The expected interest of the hedge portfolio is similar to that from an option. This is the similar assessment strategy used prior on discrete hedging for absence of transaction costs.

\item[(iv)] In non modern markets, traders are supposed to be rational, and try to increase their utility. However, if their trade activities are supposed to be rational, the decision made by the traders are explained by the two important factors. The first one refers to traders reaction to the previous stock and bond prices based on the common standardized behavior  markets. The second factor is related to the ways in which traders follow previous decisions made by the other traders. Delta-hedging strategy is one of the important components in pricing options and is utilized on the trading floor. According to the assumptions presented by Tversky and Kahneman. Following  Tversky and Kahneman's \cite{tversky} view of the availability heuristic, traders are supposed to pursue, anchor, and imitate the delta hedging Black-Scholes policy to price an option.

\end{enumerate}

\section{A pricing model for currency option in discrete time setting}\label{sec:3}
Without using the arbitrage argument, in this section we derive the pricing formula for a European currency options with transaction costs in discrete time setting. The $FBS$
equation is obtained and the sensitivity indicators are also analyzed in the latter part of this section.

For our object, a $FBS$ currency market is considered with two investment possibilities:

\begin{enumerate}
\item[(i)] A money market account:
\begin{eqnarray}
dF_t=r_dF_tdt,\qquad F_0=1,\qquad 0\leq t\leq T,
\label{eq:3}
\end{eqnarray}
where $r_d$ shows the domestic interest rate.

\item[(ii)] A stock by the following price:

\begin{eqnarray}
S_t=S_0\exp\{\mu t+\sigma \widehat{B}_H(t)\},\qquad S_0=S>0,\qquad 0\leq t\leq T,
\label{eq:4}
\end{eqnarray}
where $H>\frac{1}{2}$ is Hurst parameter.
\end{enumerate}

By using the change of variable  $B_H(t)=\frac{\mu+r_f-r_d}{\sigma}t+\widehat{B}_H(t)$, thus under the risk-neutral measure obtained:
\begin{eqnarray}
S_t=S_0\exp\{(r_d-r_f) t+\sigma B_H(t)\},\qquad S_0=S>0,\qquad 0\leq t\leq T
\label{eq:5}
\end{eqnarray}
where $r_f$ denotes foreign interest rate.\\

Let $C(t, S_t )$ be the price of a European currency option at time $t$ with a strike price $K$ that matures at time $T$. Then we
present the pricing formula for currency call option by the following theorem.

\begin{thm}
$C=C(t, S_t)$ is the value of the European call currency option on the stock $S_t$ satisfied  (\ref{eq:5}) and the trading takes place discretely with rebalancing intervals of length $\delta t$. Then $C$ satisfies the partial differential equation

\begin{eqnarray}
\frac{\partial C}{\partial t}+(r_d-r_f)S_t\frac{\partial C}{\partial S_t}+\frac{1}{2}\widehat{\sigma}^2S_t^2\frac{\partial^2C}{\partial S_t^2}-r_dC=0,
\label{eq:6}
\end{eqnarray}

and the value of the call currency option with exercise  price $K$ and expiration date $T$ is given by
\begin{eqnarray}
C=C(t,S_t)=S_te^{-r_f(T-t)} \phi(d_1)-Ke^{-r_d(T-t)}\phi(d_2).
\label{eq:7}
\end{eqnarray}

where
\begin{eqnarray}
d_1=\frac{\ln\Big(\frac{S_t}{K}\Big)+(r_d-r_f)(T-t)+\frac{\widehat{\sigma}}{2}(T-t)}{\widehat{\sigma}\sqrt{T-t}},\quad d_2=d_1-\widehat{\sigma}\sqrt{T-t},
\label{eq:9}
\end{eqnarray}
\begin{eqnarray}
\widehat{\sigma}=\sigma\Big[(\delta t)^{2H-1}+Le(H)\Big]^{\frac{1}{2}}
\label{eq:10}
\end{eqnarray}
$Le(H)=\frac{\alpha}{\sigma(\delta t)^{1-H}}\sqrt{\frac{2}{\pi}}$ is the fractional Leland number \cite{leland}and $ \phi(.)$ is the cumulative normal density function.
Moreover, using the put-call parity, we can easily obtain the valuation model for a put currency option, which is provided
by the following

\begin{eqnarray}
P=P(t,S_t)=Ke^{-r_d(T-t)}\phi(-d_2)-S_te^{-r_f(T-t)} \phi(-d_1).
\label{eq:8}
\end{eqnarray}
\label{p:2}
\end{thm}

\begin{cor}
Furthermore, if $H=\frac{1}{2}$, $\alpha=0$, from Equation (\ref{eq:6}) we have the celebrated Black-Scholes equation
\begin{eqnarray}
\frac{\partial C}{\partial t}+(r_d-r_f)S_t\frac{\partial C}{\partial S_t}+\frac{1}{2}\sigma^2S_t^2\frac{\partial^2C}{\partial S_t^2}-r_dC=0.
\label{eq:11}
\end{eqnarray}
\end{cor}

Greeks summarize how option prices change with respect to underlying variables
and are critically important in asset pricing and risk management. It can be
used to rebalance the portfolio to achieve desired exposure to a certain risk. More
importantly, knowing the Greek, a particular exposure can be hedged from adverse
changes in the market using appropriate amount of the other related financial
instruments. Unlike option prices, which can be observed in the market, Greeks can
not be observed and have to be calculated given a model assumption. Typically, the
Greeks are computed using a partial differentiation of the price formula \cite{higham,cvitanic,lyuu,shokrollahi3}.

\begin{thm} The Greeks are given by

\begin{eqnarray}
 \Delta=\frac{\partial C}{\partial S_t}=e^{-r_f(T-t)}\Phi(d_1),
 \label{eq:12}
\end{eqnarray}

\begin{eqnarray}
 \nabla=\frac{\partial C}{\partial K}=-e^{-r_d(T-t)}\Phi(d_2),
 \label{eq:13}
\end{eqnarray}

\begin{eqnarray}
 \rho_{r_d}=\frac{\partial C}{\partial r_d}=K(T-t)e^{-r_d(T-t)}\Phi(d_2),
\label{eq:14}
\end{eqnarray}

\begin{eqnarray}
 \rho_{r_f}=\frac{\partial C}{\partial r_f}=S_t(T-t)e^{-r_f(T-t)}\Phi(d_1),
 \label{eq:15}
\end{eqnarray}

\begin{eqnarray}
 \Theta&=&\frac{\partial C}{\partial t}=S_tr_fe^{-r_f(T-t)}\Phi(d_1)-Kr_de^{-r_d(T-t)}\Phi(d_2)\nonumber\\
 &-&S_te^{-r_f(T-t)}\frac{\widehat{\sigma}}{2\sqrt{T-t}}\Phi'(d_1),
 \label{eq:16}
\end{eqnarray}

\begin{eqnarray}
\Gamma=\frac{\partial^2 C}{\partial S_t^2}=e^{-r_f(T-t)}\frac{\Phi'(d_1)}{S_t\widehat{\sigma}\sqrt{T-t}},
\label{eq:17}
\end{eqnarray}

\begin{eqnarray}
\vartheta_{\widehat{\sigma}}=\frac{\partial C}{\partial \widehat{\sigma}}=S_te^{-r_f(T-t)}\sqrt{T-t}\Phi'(d_1).
\label{eq:18}
\end{eqnarray}
\label{p:3}
\end{thm}

\begin{figure}[H]
  \centering
          \includegraphics[width=1\textwidth]{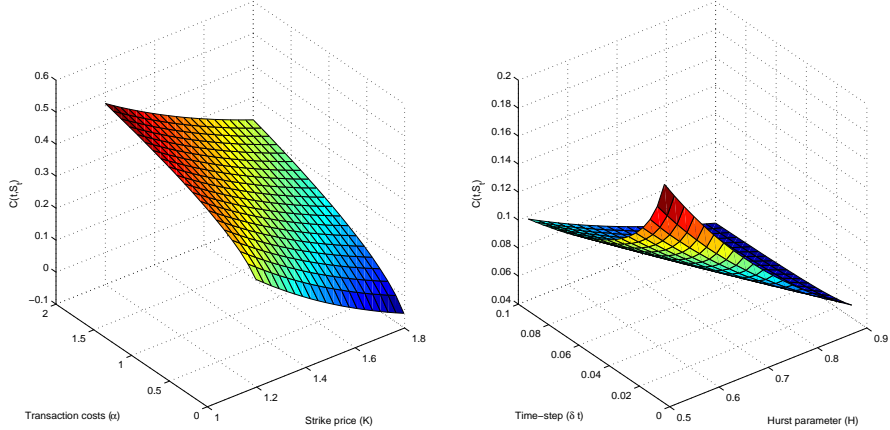}

  \caption{Impact of parameters on the $FBS$ model with transaction costs.}
\label{fig:1}
\end{figure}

It is clear that our pricing model depends on the Hurst, time-step, and transaction costs parameters. Hence we present the influence of these parameters in the following theorem and Figure \ref{fig:1}.

\begin{thm} The impact of Hurst parameter $H$, time-step $\delta t$ and transaction costs $\alpha $ are as follows

\begin{eqnarray}
\frac{\partial C}{\partial H}&=&\frac{2( \delta t)^{2H-1}\ln (\delta t)+\frac{\alpha}{\sigma}\sqrt{\frac{2}{\pi}}( \delta t)^{H-1}\ln (\delta t)}\nonumber\\
&\times&S_t\sigma^2e^{-r_f(T-t)}{2\widehat{\sigma}}\sqrt{T-t}\Phi'(d_1),
\label{eq:19}
\end{eqnarray}

\begin{eqnarray}
\frac{\partial C}{\partial \delta t}&=&\frac{(2H-1)(\delta t)^{2H-2}+\frac{\alpha}{\sigma}\sqrt{\frac{2}{\pi}}(H-1)(\delta t)^{H-2}}\nonumber\\
&\times&S_t\sigma^2e^{-r_f(T-t)}{2\widehat{\sigma}}\sqrt{T-t}\Phi'(d_1),
\label{eq:20}
\end{eqnarray}

\begin{eqnarray}
\frac{\partial C}{\partial \alpha}=\frac{S_te^{-r_f(T-t)}\sigma\sqrt{\frac{2}{\pi}}(\delta t)^{H-1}}{2\widehat{\sigma}}\sqrt{T-t}\Phi'(d_1).
\label{eq:21}
\end{eqnarray}
\label{p:4}
\end{thm}

From Figure \ref{fig:1} and Theorem \ref{p:4}, we can see that these parameters play a significant role on the $FBS$ model with transaction costs in discrete time setting.

\section{Empirical Studies}\label{sec:4}

In this section, we use the real call currency options values from the $CMB$ to assess our pricing formula. By applying the R/S method, we estimate the Hurst parameter for EUR/USD  and we achieve to $H=0.6103$. Moreover, the
estimation of volatility is obtained by considering to the historical volatility as
follows

\begin{eqnarray}
L_i=\ln\Big(\frac{q_{i+1}}{q_i}\Big),
\label{eq:25}
\end{eqnarray}

\begin{eqnarray}
\sigma=\sqrt{\frac{\sum(L_i-\overline{L})^2}{N-1}},\qquad\overline{L}=\frac{1}{N}\sum L_i,
\label{eq:26}
\end{eqnarray}
where $q_i$ shows the daily value of exchange rate.\\

These data are extracted from  the Website of $CMB$ between 01/04/2012 and 01/07/2012 (three months) with these parameters:$K=1.235, \sigma=0.1051,$ $r_d=0.0456, r_f=0.0371, T=\frac{90}{365}=0.2465, t=0.1, \delta t=0.01, \alpha=0.01$. We use the MATLAB for obtaining results by the $FBS$, mixed fractional Brownian motion $(MFBM)$ models, and the $FBS$ model with transaction costs (hereafter $TFBS$). The values calculated by different models, are indicated in Table \ref{table:1}, where $P_{Actual}$ shows the price of call currency options from $CMB$, $P_{FBS}$ denotes the values calculated by the $FBS$ model and the $P_{MFBM}$ computed the values by the $MFBM$ model and the $P_{TFBS}$ is the value computed by the $TFBS$ model. With reference to Table \ref{table:1}, it seems that the values of $FBS$, $MFBM$, and $TFBS$ models are fluctuated by the actual price from $CMB$, because the $CMB$ option values are calculated by the $BS$ model. Moreover, our results are in line with the actual price than the results obtained from the other models. In addition, values from the $TFBS$ demonstrate that whenever the time-step $\delta t$ increases, the price of call currency options will decrease. It can be said that, if we reduce the revised interval time, the pricing by our model becomes close to the actual price. This behavior is similar to the $BS$ model. These properties reveal that our $TFBS$ can get the unusual behavior from financial market and our currency pricing model seems a satisfactory model.

\begin{table}[H]
\centering
\caption{Results by different pricing models}
 \begin{tabular}{|c c c c|}
 \hline
$P_{FBS}$& $P_{MFBM}$& $P_{TFBS}$ & $P_{Actual}$\\ [0.5ex]
 \hline
0.0289& 0.0389& 0.0285 & 0.0268   \\
0.0341& 0.0455& 0.0337 & 0.0321  \\
0.0404& 0.0540& 0.0400  & 0.0372 \\
0.0594& 0.0825& 0.0590 & 0.0571    \\
0.0644& 0.0905& 0.0640 & 0.0625 \\
0.0779& 0.1126& 0.0775  &  0.0758   \\
0.0859& 0.1259& 0.0855  &  0.0836  \\
0.0929& 0.1357& 0.0925  & 0.0908  \\
0.1023& 0.1531& 0.1019  & 0.1005    \\
0.1119& 0.1688& 0.1115  & 0.1094 \\
...&...&... &...      \\ [1ex]
 \hline
\end{tabular}
\label{table:1}
\end{table}

\begin{figure}[H]
  \centering
          \includegraphics[width=1\textwidth]{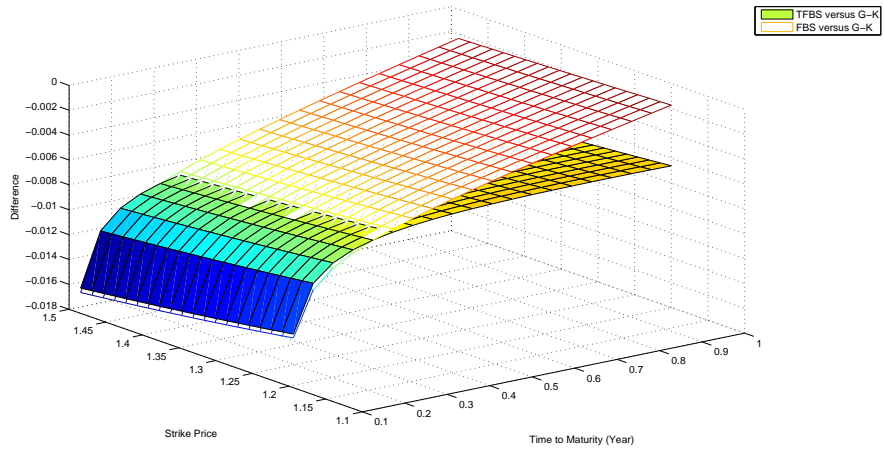}

  \caption{Relative difference among the $G-K$, $FBS$ and $TFBS$ for in-the-money case}
\label{fig:3}
\end{figure}

\begin{figure}[H]
  \centering
          \includegraphics[width=1\textwidth]{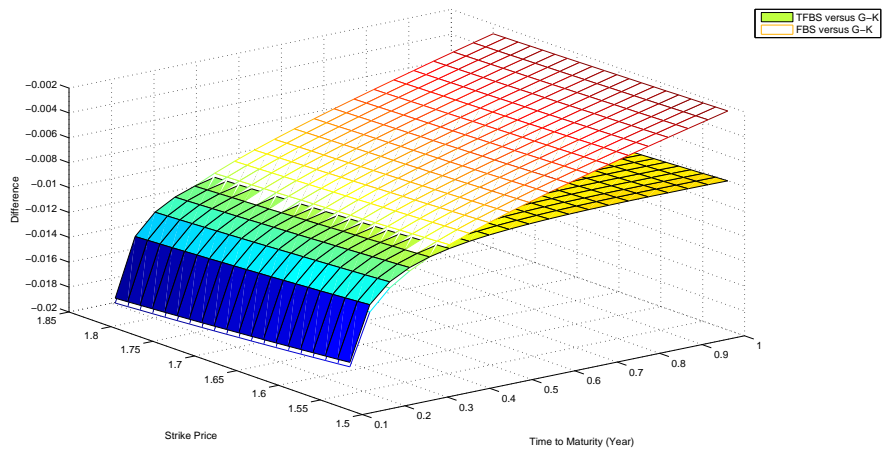}

  \caption{Relative difference among the $G-K$, $FBS$ and $TFBS$ for out-of-the-money case}
\label{fig:4}
\end{figure}

To more analyze our pricing model, we compare the prices, which are calculated by the $G-K$, $FBS$ and $TFBS$ models for both out-of-the-money and in-the-money cases. These parameters are chosen as follows: $S_t=1.512, \sigma=0.11, r_d=0.0321, r_f=0.0252, t=0.1,  \delta t=0.01, \alpha=0.1, H=0.6$ and with time maturity $T\in [0.11,0.5]$, strike price $K\in[1.2,1.49]$ for in-the-money case and $K\in[1.52,1.8]$ for out-of-the-money case. Figures \ref{fig:3} and \ref{fig:4} show the differences between the theoretical price by the $G-K$ model, $FBS$ model and our $TFBS$ model for in-the-money and out-of-the-money cases, respectively. Figures \ref{fig:3} and \ref{fig:4} show that the $TFBS$ model is better fitted with the $G-K$ model contrary to the $FBS$ model. As a result, our $TFBS$ model seems reasonable.

\section{Conclusion}\label{sec:5}
Currency options are common underlying assets that are significant derivatives in financial market. Pricing them plays an important role both in practice and theory. The present study discussed an extension European call and put currency options pricing model with transaction costs without applying the arbitrage strategy. We have displayed that the time-step $ \delta t$ and Hurst parameter $H$ are one of the significant components, in pricing currency options  with transaction costs. The estimation of volatility and Hurst parameter $H$ are also presented. Our findings showed that, since $TFBS$ model is well-developed mathematical model of huge dependence stochastic process, this model would consider as a reasonable model for pricing currency options.\\

\section*{Appendix}\label{appendix}

\textbf{Proof of Theorem \ref{p:2}.} Suppose in the replicating portfolio we have $\psi(t)$ unit of financial asset and $\varphi(t)$ unit of the riskless bond. Then, the value of the portfolio at time $t$ is
\begin{eqnarray}
P_t=\psi(t)S_t+\varphi(t)F_t.
\label{eq:27}
\end{eqnarray}

Now the movement in $S_t$ and $P_t$ is considered under discrete time interval $\delta t$. The movement in the value of the financial asset after time interval $\delta t$ is

\begin{eqnarray}
\delta S_t&&=S_t((r_d-r_f)\delta t+\sigma\delta B_H(t)+\frac{1}{2}[(r_d-r_f)\delta t+\sigma\delta B_H(t)]^2 \nonumber \\
&&+\frac{1}{6}e^{\theta[(r_d-r_f)\delta t+\sigma\delta B_H(t)]}[(r_d-r_f)\delta t+\sigma\delta B_H(t)]^3),
\label{eq:28}
\end{eqnarray}
here $\theta=\theta(t,w),\quad w\in\Omega$, and $0<\theta<1$.

Since $B_H(t)$ is continuous, from \cite{chen} we obtain

\begin{eqnarray}
(\delta t)\delta B_H(t)=O\Big((\delta t)^{1+H}\sqrt{\log (\delta t)^{-1}}\Big),
\label{eq:29}
\end{eqnarray}
\begin{eqnarray}
&&e^{\theta[(r_d-r_f)\delta t+\sigma\delta B_H(t)]}[(r_d-r_f)\delta t+\sigma\delta B_H(t)]^3\nonumber\\
&&=O((\delta t)^3)+O\Big((\delta t)^{2+H}\sqrt{\log (\delta t)^{-1}}\Big)\nonumber \\
&&+O((\delta t)^{1+2H}\log(\delta t)^{-1})+O\Big((\delta t)^{3H}(\log (\delta t)^{-1})^{\frac{3}{2}}\Big)\nonumber\\
&&=O\Big(\delta t)^{3H}(\log (\delta t)^{-1})^{\frac{3}{2}}\Big),
\label{eq:30}
\end{eqnarray}
and  $\frac{(\delta t)^{3H}(\log (\delta t)^{-1})^{\frac{3}{2}}}{(\delta t)^{1+H}(\log (\delta t)^{-1})^{\frac{1}{2}}}\rightarrow 0\quad as \quad \delta t\rightarrow 0.$\\

Then we have
\begin{eqnarray}
\delta S_t&=&(r_d-r_f)S_t\delta t+\sigma S_t\delta B_H(t)+\frac{S_t}{2}\sigma ^2(\delta B_H(t))^2\nonumber\\
&+&O\Big((\delta t)^{1+H}(\log (\delta t)^{-1})^{\frac{1}{2}}\Big),
\label{eq:31}
\end{eqnarray}

and the movement of the portfolio is
\begin{eqnarray}
\delta P_t=\psi(t)\Big(\delta S_t+r_fS_t\delta t\Big)+\varphi(t)\delta F_t-\frac{\alpha}{2}|\delta X_1(t)|S_t,
\label{eq:32}
\end{eqnarray}
where $\delta F_t$ is the movement of the money market account, $\delta \psi(t)$ is the movement of the number of units of  asset held in the portfolio.

According to the supposition (i) and \cite{leland}, transaction cost of rehedging over rehedging interval are same to $\frac{\alpha}{2}|\delta \psi(t)|S_t$.

The time interval and the asset change are both small, according to Taylor's formulae and mentioned suppositions we have
\begin{eqnarray}
\delta F_t=r_dF_t\delta t+O((\delta t)^2),
\label{eq:33}
\end{eqnarray}
\begin{eqnarray}
\delta C(t,S_t)&&=\Big(\frac{\partial C(t,S_t)}{\partial t}+(r_d-r_f)\frac{\partial C(t,S_t)}{\partial S_t}\Big)\delta t+\sigma S_t \frac{\partial C(t,S_t)}{\partial S_t}\delta B_H(t)\nonumber\\
&&+\frac{\sigma^2}{2}S_t^2\frac{\partial^2C(t,S_t)}{\partial S_t^2}(\delta B_H(t))^2+\frac{\sigma^2}{2}S_t\frac{\partial C(t,S_t)}{\partial S_t}(\delta B_H(t))^2\nonumber\\
&&+O\Big((\delta t)^{1+H}(\log(\delta t)^{-1})^{\frac{1}{2}}\Big),
\label{eq:34}
\end{eqnarray}
and
\begin{eqnarray}
\delta \psi(t,S_t)&&=\Big(\frac{\partial \psi(t)}{\partial t}+(r_d-r_f)\frac{\partial \psi(t)}{\partial S_t}\Big)\delta t+\sigma S_t \frac{\partial \psi(t)}{\partial S_t}\delta B_H(t)\nonumber\\
&&+\frac{\sigma^2}{2}S_t^2\frac{\partial^2\psi(t)}{\partial S_t^2}(\delta B_H(t))^2+\frac{\sigma^2}{2}S_t\frac{\partial \psi(t)}{\partial S_t}(\delta B_H(t))^2\nonumber\\
&&+O\Big((\delta t)^{1+H}(\log(\delta t)^{-1})^{\frac{1}{2}}\Big).
\label{eq:35}
\end{eqnarray}

From Equation (\ref{eq:35}) we obtain

\begin{eqnarray}
|\delta \psi(t,S_t)|=\sigma S_t\Big|\frac{\partial \psi(t)}{\partial S_t}\Big||\delta B_H(t)|+O(\delta t).
\label{eq:36}
\end{eqnarray}

By Equations (\ref{eq:32}), (\ref{eq:33}), (\ref{eq:36}), and $\psi=\frac{\partial C(t,S_t)}{\partial S_t}$ is obtained
\begin{eqnarray}
\delta P_t&=&r_d\varphi(t)F_t\delta t+\psi(t)(\delta S_t+r_fS_t\delta t)-\frac{\alpha \sigma}{2}S_t^2\Big|\frac{\partial \psi(t)}{\partial S_t}\Big||\delta B_H(t)|+O(\delta t)\nonumber\\
&=&\frac{\partial C(t,S_t)}{\partial S_t}\Big((r_d-r_f)S_t\delta t+\sigma S_t\delta B_H(t)+\sigma^2\frac{S_t}{2}(\delta B_H(t))^2+r_f S_t\delta t\Big)\nonumber\\
&+&r_d\Big(C(t,S_t)-S_t\frac{\partial C(t,S_t)}{\partial S_t}\Big)\delta t-\frac{\alpha \sigma}{2}S_t^2\Big|\frac{\partial^2C(t,S_t)}{\partial S_t^2}\Big||\delta B_H(t)|\label{eq:37}\\
&+&O(\delta t)\nonumber.
\end{eqnarray}

 Suppose $C=C(t,S_t)$ be replicated by the portfolio $P_t$. The value of currency option needs to same with  the value of the replicating portfolio $P_t$ to decrease (but not to eschew) arbitrage opportunities and be the stable  with economic balance.

Then \begin{eqnarray}
C(t,S_t)=\psi(t)S_t+\varphi(t)F_t  .
\label{eq:38}
\end{eqnarray}

Now we suppose that trading happen at $t$ and $t+\delta t$, but not in between which shows  the current asset price $S_t$ and the number of bonds by delta-hedging strategy held stables on the rebalancing interval $[t,t+\delta t)$.

Then, based on the suppositions (iii) and (iv), and Equations (\ref{eq:34}), (\ref{eq:37}), (\ref{eq:38}) we can get

\begin{eqnarray}
E[\delta P_t-\delta C]&=&E\Big[\Big(r_dC(t,S_t)-(r_d-r_f)S_t\frac{\partial C(t,S_t)}{\partial S_t}t\nonumber\\
&-&\frac{\partial C(t,S_t)}{\partial t}\Big)\delta -\frac{\sigma^2}{2}S_t^2\frac{\partial^2 C(t,S_t)}{\partial S_t^2}(\delta B_H(t))^2t\nonumber\\
&-&\frac{\alpha \sigma}{2}S_t^2\Big|\frac{\partial^2C(t,S_t)}{\partial S_t^2}\Big||\delta B_H(t)|+O(\delta t)\Big]\nonumber\\
&=&\Big[r_dC(t,S_t)-(r_d-r_f)S_t\frac{\partial C(t,S_t)}{\partial S_t}\nonumber\\
&-&\frac{\partial C(t,S_t)}{\partial t}-\frac{\sigma^2}{2}S_t^2(\delta t)^{2H-1}\frac{\partial^2 C(t,S_t)}{\partial S_t^2}\Big]\delta t\nonumber\\
&-&\frac{\alpha \sigma}{2}S_t^2\Big|\frac{\partial^2C(t,S_t)}{\partial S_t^2}\Big|\sqrt{\frac{2}{\pi}}(\delta t)^{H}+O(\delta t)=0,
\label{eq:39}
\end{eqnarray}
that mean self-financing delta-hedging strategy in discrete time setting.

Then

\begin{eqnarray}
&&\Big[r_dC-\Big(\frac{\partial C}{\partial t}+(r_d-r_f)S_t\frac{\partial C}{\partial S_t}+\frac{\sigma^2}{2}S_t^2(\delta t)^{2H-1}\frac{\partial^2 C}{\partial S_t^2}\nonumber\\
&&+\frac{\alpha \sigma}{2}S_t^2\sqrt{\frac{2}{\pi}}(\delta t)^{H-1}\Big|\frac{\partial^2C}{\partial S_t^2}\Big|\Big)\Big]\delta t+O(\delta t)=0.
\label{eq:40}
\end{eqnarray}
Therefore, we suppose that
\begin{eqnarray}
r_dC&=&\frac{\partial C}{\partial t}+(r_d-r_f)S_t\frac{\partial C}{\partial S_t}+\frac{\sigma^2}{2}S_t^2(\delta t)^{2H-1}\frac{\partial^2 C}{\partial S_t^2}\nonumber\\
&+&\frac{\alpha \sigma}{2}S_t^2\sqrt{\frac{2}{\pi}}(\delta t)^{H-1}\Big|\frac{\partial^2C}{\partial S_t^2}\Big|,
\label{eq:41}
\end{eqnarray}
(see \cite{leland}).
 Assume $Le(H)=\frac{\alpha}{\sigma(\delta t)^{1-H}}\sqrt{\frac{2}{\pi}}$, which is denotes fractional Leland function.

From the Equation (\ref{eq:41}) we obtain
\begin{eqnarray}
&&\frac{\partial C}{\partial t}+(r_d-r_f)S_t\frac{\partial C}{\partial S_t}+\frac{\sigma^2}{2}S_t^2(\delta t)^{2H-1}\frac{\partial^2 C}{\partial S_t^2}\nonumber\\
&&+\frac{\alpha \sigma}{2}S_t^2\sqrt{\frac{2}{\pi}}\Big|\frac{\partial^2C}{\partial S_t^2}\Big|Le(H)-r_dC=0.
\label{eq:42}
\end{eqnarray}
If $H=\frac{1}{2}$, from Equation  (\ref{eq:42}) we have
\begin{eqnarray}
&&\frac{\partial C}{\partial t}+(r_d-r_f)S_t\frac{\partial C}{\partial S_t}+\frac{\sigma^2}{2}S_t^2\frac{\partial^2 C}{\partial S_t^2}\nonumber\\
&&+\frac{\alpha \sigma}{2}S_t^2\sqrt{\frac{2}{\pi}}\Big|\frac{\partial^2C}{\partial S_t^2}\Big|Le(\frac{1}{2})-r_dC=0,
\label{eq:43}
\end{eqnarray}

that is denotes the Leland equation, and $Le(\frac{1}{2})$ is called the Leland number.

Where $\frac{\partial^2C}{\partial S_t^2}$ is ever positive for the ordinary European call option without transaction costs, if  the same conduct of $\frac{\partial^2C}{\partial S_t^2}$ is postulated here, therefore

$\Gamma$ is involved in transaction term. Equation (\ref{eq:42}) may rewrited in the form that  same  the Black-Scholes equation \cite{black}.

\begin{eqnarray}
\frac{\partial C}{\partial t}+(r_d-r_f)S_t\frac{\partial C}{\partial S_t}+\frac{\widehat{\sigma}^2}{2}S_t^2\frac{\partial^2 C}{\partial S_t^2}
-r_dC=0,
\label{eq:44}
\end{eqnarray}

where the improved volatility as follows
\begin{eqnarray}
\widehat{\sigma}=\sigma\Big[(\delta t)^{2H-1}+Le(H)\Big]^{\frac{1}{2}}.
\label{eq:45}
\end{eqnarray}

Then, from Equations (\ref{eq:44}) and (\ref{eq:45}) we can get
\begin{eqnarray}
C=C(t,S_t)=e^{-r_f(T-t)}\phi(d_1)-Ke^{-r_d(T-t)}\phi(d_2),
\label{eq:46}
\end{eqnarray}
where
\begin{eqnarray}
d_1=\frac{\ln(\frac{S_t}{K})+\Big(r_d-r_f+\frac{\widehat{\sigma}^2}{2}\Big)(T-t)}{\widehat{\sigma}\sqrt{T-t}},\quad d_2=d_1-\widehat{\sigma}\sqrt{T-t},
\label{eq:47}
\end{eqnarray}
and $\phi(.)$ is the cumulative normal distribution.

Further, if $H=\frac{1}{2}$, and $\alpha=0$, by (\ref{eq:43}) we have

\begin{eqnarray}
\frac{\partial C}{\partial t}+(r_d-r_f)S_t\frac{\partial C}{\partial S_t}+\frac{\sigma^2}{2}S_t^2\frac{\partial^2 C}{\partial S_t^2}
-r_dC=0,
\label{eq:48}
\end{eqnarray}

which is the Black-Scholes equation \cite{black}.\\

\textbf{Proof of Theorem \ref{p:3}.} First, we derive a general formula
. Let $y$ be one of the influence factors. Thus we have

\begin{eqnarray}
\frac{\partial C}{\partial y}&=&\frac{\partial
S_te^{-(r_f)(T-t)}}{\partial
y}\Phi(d_1)+S_te^{-r_f(T-t)}\frac{\partial
\Phi(d_1)}{\partial y}\nonumber\\
&-&\frac{\partial Ke^{-r_d(T-t)}}{\partial
y}\Phi(d_2)-Ke^{-r_d(T-t)}\frac{\partial \Phi(d_2)}{\partial y}.
\label{eq:49}
\end{eqnarray}

But

\begin{eqnarray}
\frac{\partial \Phi(d_2)}{\partial y}&=&\Phi'(d_2)\frac{\partial
d_2}{\partial y}\nonumber\\
&=&\frac{1}{\sqrt{2\pi}}e^{-\frac{d_2^2}{2}}\frac{\partial
d_2}{\partial y}\nonumber\\
&=&\frac{1}{\sqrt{2\pi}}\exp\Big(-\frac{(d_1-\widehat{\sigma}\sqrt{T-t})^2}{2}\Big)\frac{\partial
d_2}{\partial y}\nonumber\\
&=&\frac{1}{\sqrt{2\pi}}e^{-\frac{d_1^2}{2}}\exp\Big(d_1\widehat{\sigma}\sqrt{T-t)}\Big)\exp\Big(-\frac{\widehat{\sigma}^2(T-t)}{2}\Big)\frac{\partial
d_2}{\partial y}\nonumber\\
&=&\frac{1}{\sqrt{2\pi}}e^{-\frac{d_1^2}{2}}\exp\Big(\ln
\frac{S_t}{K}+(r_d-r_f)(T-t)\Big)\frac{\partial
d_2}{\partial y}\nonumber\\
&=&\frac{1}{\sqrt{2\pi}}e^{-\frac{d_1^2}{2}}\frac{S}{K}
\exp\Big((r_d-r_f)(T-t)\Big)\frac{\partial
d_2}{\partial y}.
\label{eq:50}
\end{eqnarray}

Then we have that
\begin{eqnarray}
\frac{\partial C}{\partial y}&=&\frac{\partial S_te^{-(r_f)(T-t)}}{\partial
y}\Phi(d_1)-\frac{\partial Ke^{-r_d(T-t)}}{\partial
y}\Phi(d_2)\nonumber\\
&+&S_te^{-r_f(T-t)}\Phi'(d_1)\frac{\partial\widehat{\sigma}\sqrt{T-t)}}{\partial y}.
\label{eq:51}
\end{eqnarray}
Substituting in (\ref{eq:51}) we get the desired Greeks.\\

\textbf{Proof of Theorem \ref{p:4}:}\\
\begin{eqnarray}
\frac{\partial C}{\partial H}&=&S_te^{-r_f(T-t)}\Phi'(d_1)\frac{\partial\widehat{\sigma}\sqrt{T-t)}}{\partial H}\Phi'(d_1)\nonumber\\
&=&S_t\sigma^2e^{-r_f(T-t)}\frac{2( \delta t)^{2H-1}\ln (\delta t)+\frac{\alpha}{\sigma}\sqrt{\frac{2}{\pi}}( \delta t)^{H-1}\ln (\delta t)}{2\widehat{\sigma}}\nonumber\\
&\times&\sqrt{T-t}\Phi'(d_1),
\label{eq:52}
\end{eqnarray}
and
\begin{eqnarray}
\frac{\partial C}{\partial \delta t}&=&S_te^{-r_f(T-t)}\Phi'(d_1)\frac{\partial\widehat{\sigma}\sqrt{T-t)}}{\partial \delta t}\Phi'(d_1)\nonumber\\
&=&S_t\sigma^2e^{-r_f(T-t)}\frac{(2H-1)(\delta t)^{2H-2}+\frac{\alpha}{\sigma}\sqrt{\frac{2}{\pi}}(H-1)(\delta t)^{H-2}}{2\widehat{\sigma}}\nonumber\\
&\times&\sqrt{T-t}\Phi'(d_1).
\label{eq:53}
\end{eqnarray}

\begin{eqnarray}
\frac{\partial C}{\partial \alpha}&=&S_te^{-r_f(T-t)}\Phi'(d_1)\frac{\partial\widehat{\sigma}\sqrt{T-t)}}{\partial \alpha}\Phi'(d_1)\nonumber\\
&=&\frac{S_te^{-r_f(T-t)}\sigma\sqrt{\frac{2}{\pi}}(\delta t)^{H-1}}{2\widehat{\sigma}}\sqrt{T-t}\Phi'(d_1).
\label{eq:54}
\end{eqnarray}

\bibliographystyle{siam}
\bibliography{../../biblio}

\begin{thebibliography}{10}

\bibitem{black}
{\sc F.~Black and M.~Scholes}, {\em The pricing of options and corporate
  liabilities}, The journal of political economy,  (1973), pp.~637--654.

\bibitem{bollen2003performance}
{\sc N.~P. Bollen and E.~Rasiel}, {\em The performance of alternative valuation
  models in the otc currency options market}, Journal of International Money
  and Finance, 22 (2003), pp.~33--64.

\bibitem{cajueiro1}
{\sc D.~O. Cajueiro and B.~M. Tabak}, {\em Testing for time-varying long-range
  dependence in volatility for emerging markets}, Physica A: Statistical
  Mechanics and its Applications, 346 (2005), pp.~577--588.

\bibitem{cajueiro2007long}
\leavevmode\vrule height 2pt depth -1.6pt width 23pt, {\em Long-range
  dependence and multifractality in the term structure of libor interest
  rates}, Physica A: Statistical Mechanics and its Applications, 373 (2007),
  pp.~603--614.

\bibitem{cajueiro}
\leavevmode\vrule height 2pt depth -1.6pt width 23pt, {\em Testing for
  time-varying long-range dependence in real state equity returns}, Chaos,
  Solitons \& Fractals, 38 (2008), pp.~293--307.

\bibitem{chen}
{\sc J.-H. Chen, F.-Y. Ren, and W.-Y. Qiu}, {\em Option pricing of a mixed
  fractional--fractional version of the black--scholes model}, Chaos, Solitons
  \& Fractals, 21 (2004), pp.~1163--1174.

\bibitem{cheridito}
{\sc P.~Cheridito}, {\em Arbitrage in fractional brownian motion models},
  Finance and Stochastics, 7 (2003), pp.~533--553.

\bibitem{cookson1992models}
{\sc R.~Cookson}, {\em Models of imperfection}, Risk, 29 (1992), pp.~55--60.

\bibitem{cvitanic}
{\sc J.~Cvitani{\'c} and F.~Zapatero}, {\em Introduction to the economics and
  mathematics of financial markets}, MIT press, 2004.

\bibitem{ding1993long}
{\sc Z.~Ding, C.~W. Granger, and R.~F. Engle}, {\em A long memory property of
  stock market returns and a new model}, Journal of empirical finance, 1
  (1993), pp.~83--106.

\bibitem{duan1999pricing}
{\sc J.-C. Duan and J.~Z. Wei}, {\em Pricing foreign currency and
  cross-currency options under garch}, The Journal of Derivatives, 7 (1999),
  pp.~51--63.

\bibitem{ekvall1997currency}
{\sc N.~Ekvall, L.~P. Jennergren, and B.~N{\"a}slund}, {\em Currency option
  pricing with mean reversion and uncovered interest parity: A revision of the
  garman-kohlhagen model}, European Journal of Operational Research, 100
  (1997), pp.~41--59.

\bibitem{garman1983foreign}
{\sc M.~B. Garman and S.~W. Kohlhagen}, {\em Foreign currency option values},
  Journal of international Money and Finance, 2 (1983), pp.~231--237.

\bibitem{higham}
{\sc D.~Higham}, {\em An introduction to financial option valuation:
  mathematics, stochastics and computation}, vol.~13, Cambridge University
  Press, 2004.

\bibitem{huang1995fractal}
{\sc B.-N. Huang and C.~W. Yang}, {\em The fractal structure in multinational
  stock returns}, Applied Economics Letters, 2 (1995), pp.~67--71.

\bibitem{leland}
{\sc H.~E. Leland}, {\em Option pricing and replication with transactions
  costs}, The journal of finance, 40 (1985), pp.~1283--1301.

\bibitem{lewellen}
{\sc J.~Lewellen}, {\em Momentum and autocorrelation in stock returns}, Review
  of Financial Studies, 15 (2002), pp.~533--564.

\bibitem{liptser}
{\sc R.~Liptser and A.~N. Shiryayev}, {\em Theory of martingales}, vol.~49,
  Springer Science \& Business Media, 2012.

\bibitem{lyuu}
{\sc Y.-D. Lyuu}, {\em Financial engineering and computation: principles,
  mathematics, algorithms}, Cambridge University Press, 2001.

\bibitem{mandelbrot}
{\sc B.~B. Mandelbrot}, {\em The fractal geometry of nature}, vol.~173,
  Macmillan, 1983.

\bibitem{mandelbrot1}
\leavevmode\vrule height 2pt depth -1.6pt width 23pt, {\em The variation of
  certain speculative prices}, Springer, 1997.

\bibitem{mariani}
{\sc M.~Mariani, I.~Florescu, M.~B. Varela, and E.~Ncheuguim}, {\em Long
  correlations and levy models applied to the study of memory effects in high
  frequency (tick) data}, Physica A: Statistical Mechanics and its
  Applications, 388 (2009), pp.~1659--1664.

\bibitem{ozdemir}
{\sc Z.~A. Ozdemir}, {\em Linkages between international stock markets: A
  multivariate long-memory approach}, Physica A: Statistical Mechanics and its
  Applications, 388 (2009), pp.~2461--2468.

\bibitem{podobnik2008detrended}
{\sc B.~Podobnik and H.~E. Stanley}, {\em Detrended cross-correlation analysis:
  a new method for analyzing two nonstationary time series}, Physical review
  letters, 100 (2008), p.~084102.

\bibitem{rogers}
{\sc L.~C.~G. Rogers et~al.}, {\em Arbitrage with fractional brownian motion},
  Mathematical Finance, 7 (1997), pp.~95--105.

\bibitem{rosenberg1998pricing}
{\sc J.~V. Rosenberg}, {\em Pricing multivariate contingent claims using
  estimated risk--neutral density functions}, Journal of International Money
  and Finance, 17 (1998), pp.~229--247.

\bibitem{salopek}
{\sc D.~M. Salopek}, {\em Tolerance to arbitrage}, Stochastic Processes and
  their Applications, 76 (1998), pp.~217--230.

\bibitem{sarwar2000empirical}
{\sc G.~Sarwar and T.~Krehbiel}, {\em Empirical performance of alternative
  pricing models of currency options}, Journal of Futures Markets, 20 (2000),
  pp.~265--291.

\bibitem{shokrollahi1}
{\sc F.~Shokrollahi and A.~K{\i}l{\i}{\c{c}}man}, {\em Pricing currency option
  in a mixed fractional {B}rownian motion with jumps environment}, Math. Probl.
  Eng.,  (2014), pp.~Art. ID 858210, 13.

\bibitem{shokrollahi4}
\leavevmode\vrule height 2pt depth -1.6pt width 23pt, {\em Actuarial approach
  in a mixed fractional {B}rownian motion with jumps environment for pricing
  currency option}, Adv. Difference Equ.,  (2015), pp.~2015:257, 8.

\bibitem{shokrollahi5}
\leavevmode\vrule height 2pt depth -1.6pt width 23pt, {\em The valuation of
  currency options by fractional brownian motion}, SpringerPlus, 5 (2016),
  p.~1145.

\bibitem{shokrollahi3}
{\sc F.~Shokrollahi, A.~K{\i}l{\i}{\c{c}}man, N.~A. Ibrahim, and F.~Ismail},
  {\em Greeks and partial differential equations for some pricing currency
  options models}, Malays. J. Math. Sci., 9 (2015), pp.~417--442.

\bibitem{shokrollahi2}
{\sc F.~Shokrollahi, A.~K{\i}l{\i}{\c{c}}man, and M.~Magdziarz}, {\em Pricing
  {E}uropean options and currency options by time changed mixed fractional
  {B}rownian motion with transaction costs}, Int. J. Financ. Eng., 3 (2016),
  pp.~1650003, 22.

\bibitem{tversky}
{\sc A.~Tversky and D.~Kahneman}, {\em Availability: A heuristic for judging
  frequency and probability}, Cognitive psychology, 5 (1973), pp.~207--232.

\bibitem{wang2010scaling0}
{\sc X.-T. Wang}, {\em Scaling and long-range dependence in option pricing i:
  pricing european option with transaction costs under the fractional
  black--scholes model}, Physica A: Statistical Mechanics and its Applications,
  389 (2010), pp.~438--444.

\bibitem{wang2010scaling}
{\sc X.-T. Wang, E.-H. Zhu, M.-M. Tang, and H.-G. Yan}, {\em Scaling and
  long-range dependence in option pricing ii: Pricing european option with
  transaction costs under the mixed brownian--fractional brownian model},
  Physica A: Statistical Mechanics and its Applications, 389 (2010),
  pp.~445--451.

\end{thebibliography}

\end{document}